\documentclass[aps,prx,twocolumn,superscriptaddress,showpacs]{revtex4-2}
\usepackage[normalem]{ulem}
\usepackage{amssymb}
\usepackage{suffix}
\usepackage{mathtools}
\usepackage[utf8]{inputenc}
\usepackage{booktabs}
\usepackage{cases}
\usepackage[multiple]{footmisc}
\usepackage{dcolumn}
\usepackage{color,soul}
\usepackage{rotating}
\usepackage{perpage}
\usepackage{xcolor}
\usepackage{soul}
\usepackage{tikz}
\usepackage[T1]{fontenc}
\usepackage{etoolbox}
\usepackage{graphics}
\usepackage{siunitx}
\usepackage{hyperref}
\usepackage{float}
\usepackage{multirow}
\usepackage{mathtools}
\usepackage{bm}
\usepackage{url}

\usepackage{tikz}
\usepackage{tikz-3dplot}
\usepackage [outdir=./]{epstopdf}

\begin{document}
\title{Light-induced topological phases in thin films of magnetically doped topological insulators}

\author{S. Sajad Dabiri}
\address{Department of Physics, Shahid Beheshti University, 1983969411 Tehran, Iran}
\author{Hosein Cheraghchi}
\email{cheraghchi@du.ac.ir}
\address{School of Physics, Damghan University, P.O. Box 36716-41167, Damghan, Iran}
\address{School of Physics, Institute for Research in Fundamental Sciences (IPM), 19395-5531, Tehran, Iran}

\author{Ali Sadeghi}
\email{ali_sadeghi@sbu.ac.ir}
\address{Department of Physics, Shahid Beheshti University, 1983969411 Tehran, Iran}
\address{School of Nano Science, Institute for Research in Fundamental Sciences (IPM), 19395-5531, Tehran, Iran}

\date{\today}
\vspace{1cm}
\newbox\absbox
\begin{abstract}
We study the photon-dressed electronic band structure of topological insulator 
thin films 
doped by magnetic impurities
in response to an off-resonance time-periodic electromagnetic field. 
The  thin films   irradiated by a circularly polarized light
undergo phase transition, 
and a fascinating feature of distinct phases emerges in the phase diagram 
depending on the 
frequency, intensity and polarization of the light.
As a particular case, quantum anomalous Hall insulator 
phase is induced purely by the light-induced mass term with no need to any external magnetic field or even magnetization arising from 
doping magnetic impurities. 
Moreover, a novel phase, \emph{quantum pseudo-spin Hall insulator (QPHI)}, emerges in the phase diagram leading to \emph{anisotropic helical edge} states with zero total Chern number. 
We verify these  achievements 
by numerical calculations for a nanoribbon of the thin film for which the edge mode behavior is observed 
at several points on the phase diagram. 
The emergence of the mentioned topological phases and the edge modes are further confirmed 
by both calculating the Hall conductivity by means of the Kubo formula and the Chern number of each band. The effect of light parameters on the Landau level fan diagram in the presence of a perpendicular magnetic field indicates various topological phases occurring at higher Chern numbers.
\end{abstract}
\maketitle

\section{Introduction}
Quantum anomalous Hall effect
which was firstly observed in a thin film of the Cr-doped (Bi,Sb)$_2$Te$_3$ family materials~\cite{science329,science340,prl113137201,nature10731},
is the variant of the quantum Hall effect without an external magnetic field.
Magnetization (induced by magnetically doped impurities) and structural inversion asymmetry (SIA) 
can change the topological invariant of a topological insulator (TI) thin film~\cite{electrically,universality,prb81H,prl2013QT,science340,zeemanelec}.A perpendicular magnetic field can also lead to emergence of Landau levels and phases with higher Chern numbers \cite{zyuzin,edge,feng,QHETI}.
Therefore, taking these parameters into account gives us a static degree of freedom to engineer the topological phases of these materials.
On the other hand, recent technological progresses in mid-infrared lasers  has established a new route for 
engineering the electronic band structure. 
Indeed, an interplay between the periodicity of lattice and time coming from 
oscillating field
extends the Hilbert space by inducing the so-called Floquet-Bloch states
and enables one to modify the band structure at will by using photon-assisted processes~\cite{2013Gomez,exp_haldane}. 
The Floquet-Bloch states have been experimentally observed in the irradiated topological insulator $\text{Bi}_2\text{Se}_3$
where
the angle-resolved photo-emission spectroscopy (ARPES) of the gap-less surface states show replicas of the Dirac cones 
separated by photo-induced gaps~\cite{2013Wang}. 
The dynamical gaps emerging in the band spectrum depend on the radiation intensity and polarization.
Modification of the band spectrum has also been studied in ultra-cold atom gases~\cite{2}, graphene structures~\cite{2011Gu,2014Perez,2013Delplace}, HgTe/CdTe quantum wells~\cite{2013Cayssol,2011Lindner}, etc. 
Furthermore, Floquet topological insulator is a new state with non-trivial topology which is realized by illumination of a mid-infrared laser on materials~\cite{2011Lindner,2010Kitagawa}. 
To extend such a new quantum state induced by illumination, topological protected edge states emerge not only at the edge of irradiated graphene sheets but also around large defects and adatoms at the dynamical gap~\cite{2016Lovey}. 

Circularly polarized light has been shown to affect the mass term of the Hamiltonian and, as a consequence, the Berry curvature of the driven system which in turn gives rise to the emergence of non-trivial topological states in the material~\cite{2010Kitagawa}. 
Interestingly, in driven two dimensional systems, two robust co-propagating chiral edge modes, the so-called ``anomalous edge states'', appear even though the Chern numbers of bulk Floquet bands are zero. The appropriate definition of the Chern number for this new phenomenon  is the difference between the number of edge states inside the top and bottom dynamical gaps which encompass the given bulk Floquet bands\cite{rudner}. These dynamical gaps could be hosted a large
number of co-propagating edge modes if, as induced by system parameters, 
a band touching occurs and two Floquet band exchange their Chern numbers\cite{creat_edge}. However, for driven frequencies much larger than the band width, the Floquet eigen-states are mostly localized on the central Floquet band leading to an effective driven Hamiltonian. In this way, a TI induced by time-periodic perturbations is achievable from an originally undriven semi-metal or normal insulator. The  phase transition might also happen between two distinct TI phases with exotic edge modes such as pseudo-spin modes.

In this manuscript,  we study the perturbative effect of the irradiation of a laser with circular polarization on the Hamiltonian in the off-resonant regime 
and investigate the new topological phases appeared in magnetically doped TI thin films. 
We demonstrate that a left- or right-handed circularly polarized laser introduces an additional mass term into the 
Hamiltonian leading to a rich phase diagram. 
 The tunneling between the opposite surfaces of the film opens a gap in the spectrum
 and the tunneling nature has a crucial role in determining the light-induced phases. 
We show that depending on the film thickness 
the  phase diagram 
can contain a phase transition from quantum anomalous Hall insulator (QAHI)
to normal insulator (NI) and vice versa.
In addition to QAHI, a new quantum phase emerges where the film behaves like a quantum spin Hall insulator but its edge modes contain the pseudo-spin states and we therefore call it ``quantum pseudo-spin Hall insulator'' (QPHI). 
By tuning the intensity and polarization of the radiation, one can engineer new quantum states of matter with unique dispersions. Finally, we will show that if the TI thin film is subjected simultaneously to both magnetic field and circularly polarized light, the Landau-level fan diagrams split into a fascinating structure of phases showcasing higher Chern number states.  
Thanks to the easy access controllability of light intensity and frequency, the light-induced manipulation of new quantum states of the matter which is hard to fabricate on materials would be worthwhile.

The rest of this manuscript is organized as follows. 
In Sec.~\ref{S2}, we review the effective low-energy Hamiltonian in the absence (dark) or 
presence (photo-induced)  of light.
We then introduce our modified Hamiltonian for the special case of perpendicular magnetization in response to radiation.
The first part of Sec.~\ref{S3} is devoted to presenting and discussing the 
emergent phases on the phase diagram.
We then calculate the Hall conductivity within the Kubo formalism 
followed 
by searching for the edge modes in the spectrum by discretizing the photo-induced Hamiltonian on a  
nanoribbon . 
We also present and discuss the Landau level fan diagrams 
 before we draw our conclusions in Sec.~\ref{S5}.

\section{Effective low-energy Hamiltonian}\label{S2}
\subsection{Dark Hamiltonian}\label{S2_a}
 In an ultra-thin film of  TI materials such as the (Bi,Sb)$_2$Te$_3$ family,
two Dirac cones, 
associated to the helical surface states on the top and bottom surfaces,
are hybridized with each other.
As a result, tunneling between the two surface states 
opens a gap in the band structure.
On the other hand, a strong perpendicular ferromagnetism emerges in this thin film arising from the interaction between doped magnetic impurities. 
The resultant magnetization is assumed as a new control parameter of the gap size and the new quantum phase is 
QAHI.

The effective low-energy Hamiltonian  for the   hybridized Dirac cones near the $\Gamma$ point is given by~\cite{effective}
\begin{equation}
\begin{aligned}
\centering
H^{\text{dark}}(\bf{k})
&= \hbar {v}_f (k_y \sigma _x -k_x \sigma _y) \otimes \tau_z
+\Delta (\textbf{k}) \sigma _0 \otimes \tau _x
\\&+V_\text{sia} \sigma _0 \otimes \tau _z
+ M_z \sigma_z\otimes\tau _0.
\end{aligned}
\label{eq:firsthamil}
\end{equation}
Here $\sigma _0$ and $\tau_0$ are 2$\times$2 identity matrices
while $\tau_i$ and $\sigma_i$ ($i=x,y,z$) are the Pauli matrices represented in the basis sets of 
    the thin film upper and lower surfaces states $\{|u\rangle,|d\rangle\}$ and 
    the electron spin states $\{|\uparrow \rangle, |\downarrow\rangle\}$, respectively.  
The mass term $\Delta$, induced by tunneling between the top and bottom surfaces,
is experimentally  mapped on to $\Delta(\textbf{k})=\Delta_0+\Delta_1 k^2$ for films of Bi$_2$Se$_3$ and the (Bi,Sb)$_2$Te$_3$ family thinner than $\sim5~\text{nm}$.~\cite{nature584,prb81041307} 
In this work, in our numerical calculations, we consider system parameters as $\Delta_0=35$~meV and $\Delta_1=\pm10$~eV\AA$^2$
and  set the Fermi velocity to $v_f=4.48 \times 10^{5}$~m/s. To guarantee the off-resonant regime in which the central Floquet band is far away other replicas, we set driven frequency as $\hbar \Omega=1$~eV which is much larger than the band width.

 The SIA potential, $V_\text{sia}$, appeared in the third term is caused by a kind of perpendicular potential difference induced by the applied electric field or also by the substrate~\cite{nature584}  giving rise to a Rashba-like splitting of the band spectrum. Indeed, the SIA potential shifts one of the two surface Dirac cones upward and the other one downward on the energy axis. The asymmetry impact of the substrate is weaker in thinner films. Also, this term may
 stem from the difference between the environment of the top and bottom surfaces, which is stronger for thinner films~\cite{nature584}.
 The last term in Eq.~(\ref{eq:firsthamil}) refers to the perpendicular strong ferromagnetic exchange field 
 which originates from the magnetic impurities such as Ti, V, Cr and Fe doped in Bi$_2$Se$_3$ and (Bi,Sb)$_2$ Te$_3$ family~\cite{Magnetic}.

Equation~(\ref{eq:firsthamil}) expresses the Hamiltonian in the spin-surface Hilbert space spanned by the basis set $\{|u,\uparrow \rangle ,|u,\downarrow \rangle ,|d,\uparrow \rangle ,|d,\downarrow \rangle\}$.
In some occasions, it is more convenient to  apply a unitary transformation to the surface states 
and work with the bonding and anti-bonding states
$|\psi_b     \rangle = \left( | u\rangle + | d\rangle \right)/ \sqrt 2  ,
|\psi_{ab}  \rangle =  \left( | u\rangle - | d\rangle \right)/ \sqrt 2 $.
In terms of the basis set 
$\{|\psi_b,\uparrow \rangle ,|\psi_{ab},\downarrow \rangle ,|\psi_b,\downarrow \rangle ,|\psi_{ab},\uparrow \rangle\}$, 
the Hamiltonian is expressed as 
\begin{equation}
\centering
H^{\text{dark}} (\textbf{k})=\left(\begin{array}{cc} h_{+} (\textbf{k}) & V_\text{sia}\sigma_x\\ V_\text{sia} \sigma_x& h_{-}(\textbf{k}) \\\end{array} \right)
.
\label{eq:hamiltotalchiral} 
\end{equation}
In the particular case $V_\text{sia}=0$,  even for $M_z \neq 0$ the effective Hamiltonian 
reduces to a two-block diagonal matrix
\newcommand{\az}{{\alpha_z}}
whose blocks are characterized by the pseudo-spin index $\az=+,-$ as~\cite{universality,geometrical,PRL2013_zhang} 
\begin{equation}
\centering
h_\az^{\text{dark}(\textbf{k})}=\hbar v_f \left(k_y \sigma _x- \az k_x \sigma _y\right)+\Big(\Delta(\textbf{k})+\az M_z \Big)\sigma _z
\label{eq:hamilchiral} 
 \end{equation}
 each of which has its own pseudo-spin band spectrum 
\begin{equation}
\centering
E_\az(k)=\pm \sqrt{\left(\hbar v_fk \right)^2+\left(\Delta_0 +\Delta_1 k^2+\az M_z\right)^2}
\label{eq:Echiral} 
.\end{equation}

\subsection{Photo-induced Hamiltonian} \label{S2_b}
Let us now consider a high frequency electromagnetic field with circular polarization irradiated on a thin film 
TI sample of magnetically doped Bi$_2$Se$_3$ or (Bi,Sb)$_2$Te$_3$ family. 
The vector potential representing the magnetic component of the radiation is given by
\begin{equation}
{\textbf A}(t)=A_0(\sin\Omega t,\cos\Omega t)
\end{equation} 
where $\Omega=2 \pi /T$ is the frequency of light so that 
${\textbf A}(t+T)={\textbf A}(t)$
while $A_0$ is adjustable in experiment by simply tuning the light intensity. 
Here, $\Omega>0$ is attributed to the right-hand circularly polarized light while $\Omega<0$ refers to a left-handed circular polarization.
Radiation can be introduced into the Hamiltonian 
by replacing the wave-vector ${\textbf k}$ by ${\textbf k}+e {\textbf A}/\hbar$,

\begin{equation}
 H(t)=H^{\text{dark}}(\textbf{k}+e {\textbf A}/\hbar). 
\end{equation} 
We work in the off-resonant regime
where $\Omega$ is much larger than the band width of the system. 
Indeed, the band width of the surface band structure is, at most, equal to the band gap of bulk Bi$_2$Se$_3$ or (Bi,Sb)$_2$Te$_3$ family, i.e. $\sim$0.3~eV. Therefore, taking 1~eV for the laser beam energy 
guarantees that the replicas of the bulk Floquet bands split up and, by varying the system parameters,  there is no band touching of other Floquet side bands with the bulk Floquet band at zero energy. Let us stress that in this work, we are focusing on the edge modes falling inside the band gap of the first bulk Floquet band, not on the edge modes appeared in the dynamical gaps. Moreover, the irradiated photon does not excite an electron directly but 
its effect is described by a time-averaged Hamiltonian.
For a very weak laser field of scaled strength $\mathcal{A} \equiv {e A_0}/{\hbar} \ll 1/a_0$, where $a_0\simeq 4$~\AA~ is the lattice parameter of (Bi,Sb)$_2$Te$_3$ family materials \cite{nature2009},
 the effective Hamiltonian for high-frequency regime is expanded as~\cite{Bukov,2010Kitagawa,ezawasilicene}
\begin{equation}
\mathcal{H}=\mathcal{H}^0+\sum_{j=1}^\infty{\frac {\left[\mathcal{H}^{-j},\mathcal{H}^{+j}\right]} {j\hbar \Omega}}+\mathcal{O}(\frac{1}{(\hbar \Omega)^{2}})
\label{one_photon_H}
\end{equation}

Here 
${\mathcal{H}} ^ {n}= \frac 1 T \int_{0}^{T} \mathcal{H}(t) e^{ i n|\Omega| t} dt$ where $n=0,\pm1,...$ 
corresponds to virtually n-photon absorption/emission
so that 
\[ h_\az^0=h_\az^{\text{dark}}+\Delta_1 \mathcal{A}^2 \sigma _z
\]
and the one-photon assisted version of Eq.~(\ref{eq:hamilchiral}) reads
\[h^{\pm1}_\az=\frac 1 2 \hbar v_f \mathcal{A} \left(\sigma_x \mp i \az \sigma_y\right)+\Delta_1 \mathcal{A} 
\left(\pm ik_x+ k_y\right) \sigma_z 
.\] 
The higher photon assisted versions are equal to zero.
The commutation between the two decoupled 
pseudo-spin Hamiltonians, namely

\[
\frac 1 {\hbar \Omega}\left[h^{-1}_\az,h^{+1}_\az\right] =
 \frac  {2v_f \mathcal{A}^2}{\Omega} \Delta_1(k_x \sigma_y-\az k_y \sigma_x)
+\az m_{\Omega} \sigma_z \]
leads to the effective static Hamiltonian 
\newcommand{\etaz}{\eta_\az}
\newcommand{\deltaz}{\delta_\az}
\begin{equation}
h^{driven}_{\az}=\hbar \etaz v_f ( k_y \sigma_x-\az k_x \sigma_y)+[\Delta^\prime({\bf k})+\az(M_z+m_{\Omega})]\sigma_z
\label{Photo_Hamiltonian}
\end{equation}
with the definition $\Delta^\prime({\bf k})=\Delta_0^\prime+\Delta_1 k^2$ and $\Delta_0^\prime=\Delta_0+\Delta_1\mathcal{A}^2$. The band spectrum would be 
\begin{equation}
E_\az(k)=\pm \sqrt{( \etaz \hbar v_f k)^2+( \deltaz+\Delta_1 k^2)^2}
\label{chiral_photospectrum}
 \end{equation}
 for each of the pseudo-spin indices $\az$. 
The mass term 
$m_{\Omega}={(\hbar v_f \mathcal{A})^2}/{\hbar\Omega}$ 
induced by light irradiation
is grouped by the one in Eq.~(\ref{eq:hamilchiral}) into the modified total mass term 
\begin{equation} 
\deltaz=\Delta_0^\prime+\az(M_z+m_{\Omega})
\label{eq:delta}
.\end{equation}
Interestingly, compared to the dark case described by Eqs.~(\ref{eq:hamilchiral}) and (\ref{eq:Echiral}), 
$v_f$ is now scaled by  a factor 
\begin{equation}
\etaz=1-\frac{ 2\az \mathcal{A}^2 }{\hbar \Omega} \Delta_1
\label{eq:velocity}
.\end{equation}
This implies that the Fermi velocity of helical surface states is modulated depending on the light parameters.
This is a hallmark of new topological phases in the irradiated material. 

\subsection{Topological Invariants} \label{sec:TInvar}
For a Hamiltonian which commutes with the \emph{pseudo-spin operator} $\hat \az$
and thus is block diagonal in the pseudo-spin space,
the band spectrum splits into two $\pm$ pseudo-spin bands
with pseudo-spin Chern number $C_\az$ per each block \cite{Prodan,Sheng,spinchernnumber,spinchernnumberTI}. 
One defines the total Chern number as $C=C_{+}+C_{-}$ and the total pseudo-spin Chern number as $C'= \frac {1}{2} (C_+-C_-)$.

By integrating the Berry curvature 
\begin{equation}
F_{\az}= \frac{-\az}{4\pi} 
\frac{(\hbar\eta_{\az}v_f)^2(\delta_{\az}-\Delta_1k^2)}
{\big(k^2(\hbar\eta_{\az}v_f)^2+(\delta_{\az}+\Delta_1k^2)^2\big)^{3/2}}
\label{lll}
\end{equation}
 over the whole Brillouin zone, one gains the Chern number for each pseudo-spin sector
\begin{equation}
C_\az=\frac \az 2  \big (\text{sgn}\Delta_1 -\text{sgn} \deltaz \big)
\label{pseudo-spinchernnumber}
.\end{equation}
It is interesting that the latter does not depend on $\eta_{\alpha_z}$. Although Eq. \ref{pseudo-spinchernnumber} only works if the pseudo-spin is conserved, pseudo-spin Chern number can be still defined in a tricky complicated way even if $\az$ is not conserved ~\cite{Prodan,Sheng}.This definition is generally based on the smooth decomposition of the occupied valence band into two sectors during the digonalization of pseudo-spin operator $\hat \az$ in the valence
band. This quantity has been calculated in Ref. \cite{spinchernnumberTI} explicitly for TI thin films when there is no pseudo-spin conservation. Since we set $V_{sia}$ to be zero, pseudo-spin is conserved in this work.

If time reversal symmetry is present (corresponding to $M_z=m_\Omega=0$ in our model),
$C' \bmod 2$ and the $\mathbb{Z}_2$ index \cite{prl95Z} yield the same result and both are topological invariants of the system. Such a relationship is throughly investigated for TI thin films in Ref. \cite{spinchernnumberTI}.
However, even if there is no time reversal symmetry so that the $\mathbb{Z}_2$ invariant is not well-defined, 
one can still use $C_\az$.  
Nevertheless, $C_\az$ cannot protect the edge modes when both time reversal symmetry and pseudo-spin conservation are broken.

\section{Results and Discussion} \label{S3}

\subsection{Phase Diagram}\label{sec:phasediag}

\begin{figure*}
\includegraphics[width=0.95\linewidth,trim={0 0 0.25cm 0}, clip]{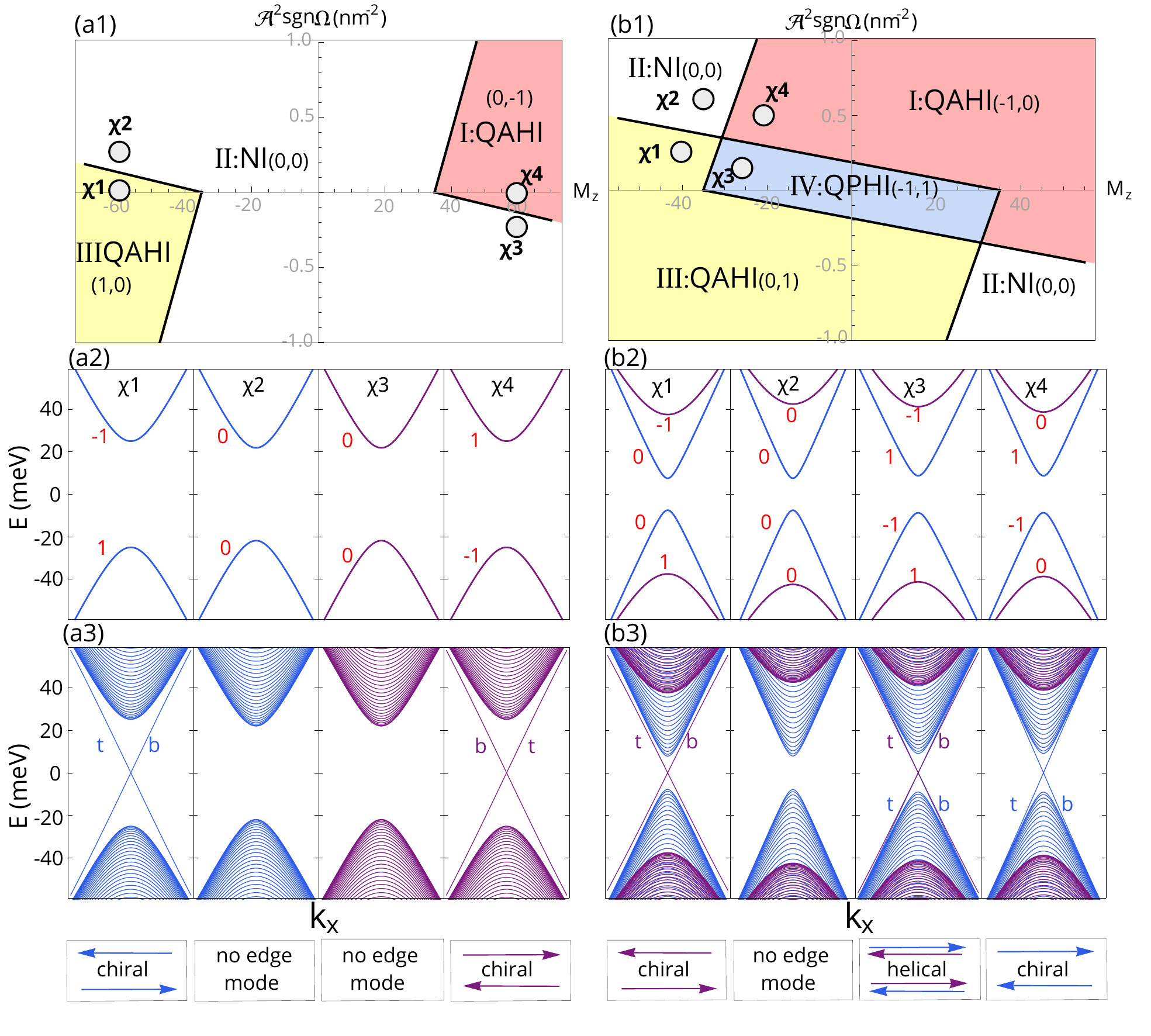}

\caption{Phase diagram induced by light illumination and the band structure for selected phases
for case (i), ${\Delta_0}{\Delta_1}>0$, (left) and case (ii), ${\Delta_0}{\Delta_1}<0$, (right). 
The special points $\chi1$ to $\chi4$ in the phase diagrams (a1) correspond to
($M_z,\mathcal{A}^2\text{sgn}{\Omega}$) for the special points $\chi1$ to $\chi4$ correspond to
(-60,0), (-60,0.25), (60,-0.25) and (60,0) in (a1), and to (-40,0.2), (-35,0.6), (-25,0.1) and (-20,0.5) in (b1) in units of (meV,1/nm$^2$).
The bands corresponding to these phases are shown in panels (a2) and (b2)
while the energy bands in (a3) and (b3) correspond to a square-lattice nanoribbon of the same TI thin film of lattice parameter 2~nm and width 400~nm. 
The blue and purple colors in the spectra refer to  the pseudo-spin indices $\az=+1$ and $\az=-1$, respectively,
while ``t'' and ``b'' to the top and bottom edges. The lower panels are a schematic representation of the edge modes and currents.}
\label{phase_diagram}
\end{figure*}

\begin{figure*}   
 \includegraphics[width=0.9\linewidth] {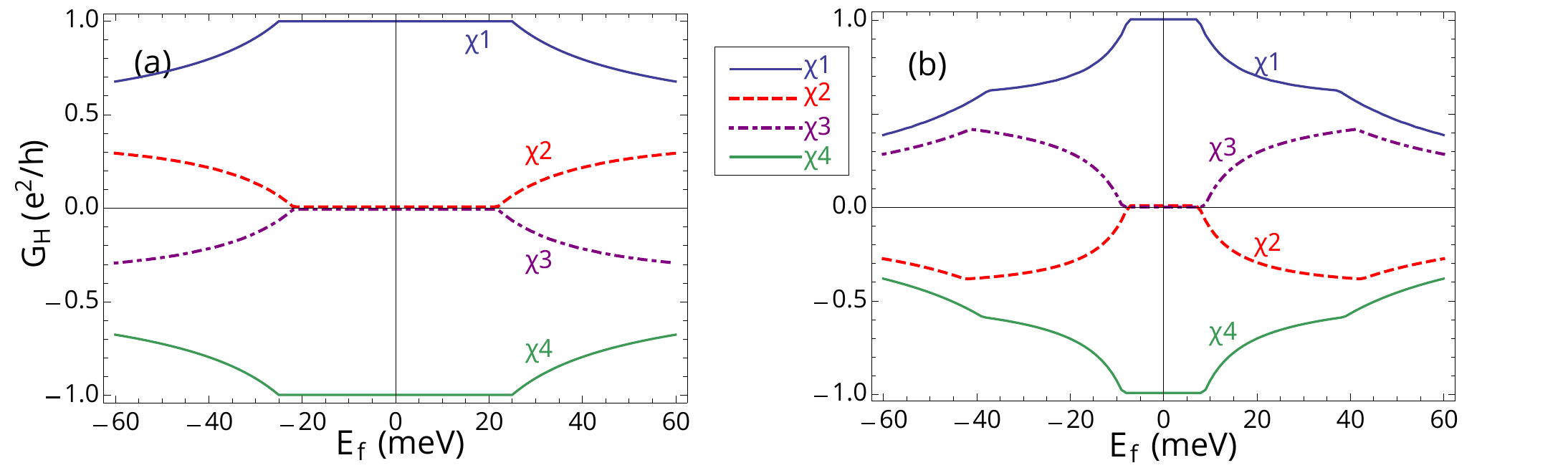}    
 \caption{Zero temperature Hall conductivity, as calculated by Kubo formula, Eq.~(\ref{Kubo}),
 for four special states $\chi1$-$\chi4$, in the case of ${\Delta_0}{\Delta_1}>0$ (left) or ${\Delta_0}{\Delta_1}<0$ (right).}
\label{kubofig}
\end{figure*}

In the following, we first investigate the effect of radiation of a weak laser on the electronic phases of 
a thin film TI sample without an in-plane magnetization nor a SIA potential. 
The spectrum of the effective static Hamiltonian, 
reduced to two decoupled pseudo-spin Hamiltonians given by Eq.~(\ref{chiral_photospectrum}),
 shows several interesting non-trivial electronic states depending on the experimentally tunable parameters.
 In particular, the Fermi velocity $v_f$ is effectively modified by the radiation frequency and intensity while 
the two parameters $\Delta_0$ and $\Delta_1$ depend on the film thickness.
We present the results for two  distinct cases, (i) ${\Delta_0}{\Delta_1}>0$ and (ii) ${\Delta_0}{\Delta_1}<0$,
on the left and right columns in Fig.~\ref{phase_diagram}, respectively.
For simplicity and without loss of generality, we assume that $\Delta_0>0$ so that the two cases are determined merely by  sgn$\Delta_1$.

In practice, by changing the intensity, frequency and polarization of the light
one can tune $m_{\Omega}$ and engineer the topological phase of the irradiated thin film. 
Let us start from the  solid lines $m_\Omega=-M_z\pm \Delta_0^\prime$,
corresponding to  $\delta_\pm=0$ in Eq.~(\ref{eq:delta}),
which divide the  $M_z - \mathcal{A}^2$ plane into several topologically different regions
denoted by I, II, III and IV
on the phase diagrams illustrated in Figs.~\ref{phase_diagram}~(a1,b1).
The band structures of four illustrative points $\chi1, \chi2, \chi3$ and $\chi4$ shown on panels (a2,b2)
are meant to represent the smooth evolution of the band structure from one region  to the next when
 $M_z$ and $\mathcal{A}^2$ vary.
Note that the bands assigned to pseudo-spin index $\az=+ (-)$ are identified by blue (purple) color.
Band inversion occurs across a boundary on which the Chern number undergoes a change.
A first set of the represented gapless states lie on the borders between the regions.
Indeed, Eq.~(\ref{chiral_photospectrum}) suggests that 
the spectrum has no gap at the $\Gamma$-point ($k=0$) if the total mass term  $\deltaz$ is nullified 
for either of the pseudo-spin bands i.e.  along the boundaries: 
If the light-induced mass term is tuned to $m_{\Omega}=-M_z-\Delta_0^\prime$ so that $\delta_+=0$,
a single Dirac cone spectrum (SDC)  with $E_+(k)\simeq \pm \eta_+ \hbar v_f k$
is formed around the $\Gamma$-point.
Similarly, $m_\Omega=-M_z+\Delta_0^\prime$ leads to another $\Gamma$-centered SDC band touching $E_-(k)\simeq \pm  \eta_- \hbar v_f k$  for the other pseudo-spin index.
Region (II) and (IV) are delimited by the linear boundaries, i.e. ${|M_z+m_{\Omega}|<|\Delta_0^\prime|}$.
 thus from Eq. \ref{pseudo-spinchernnumber}
\begin{equation}
\small (C_+,C_-) =
\begin{cases}
(0,0) & \text{if } \Delta_0^\prime \Delta_1>0  
\\
(-1,+1) &\text{if } \Delta_0^\prime \Delta_1<0  
.\end{cases}
\end{equation}
For case (i), i.e. $\Delta_1>0$,  
the total pseudo-spin Chern number is zero and we have a NI phase in region (II) in panel (a1) of Fig. \ref{phase_diagram}. 
 On the other hand, if $\Delta_1<0$, the total Chern number is again zero 
 but, there are a NI phase in  region (II) and QPHI phase in region (IV). In QPHI phase, like the  QSHI phase, the helical edge states that are composed of pseudo-spin states flow anisotropically along the edges (panel (b1) of Fig. \ref{phase_diagram}).

In \emph{regions (I) and (III)} where ${|M_z+m_{\Omega}|>|\Delta_0^\prime|}$, 
an inverted band spectrum with a non-trivial band topology is observed.
For case (i), i.e. $\Delta_1>0$, Eq.~(\ref{pseudo-spinchernnumber}) leads to 
\begin{equation}
\small (C_+,C_-) =
\begin{cases}
(0,-1) &\text{if } M_z+m_{\Omega}>0 
\\
(+1,0) &\text{if } M_z+m_{\Omega}<0 
\end{cases}
.\end{equation}
For case (ii) ($\Delta_1<0$) one exchanges the values of $C_+$ and $C_-$.
It should be noted that 
the total Chern number $C=-(M_z+m_{\Omega})/|M_z+m_{\Omega}|$ is non-zero which is an indication of the QAHI phase
in both cases.

Although  a band gap emerges at the $\Gamma$ point for states away from the boundaries between the regions,
Eq.~(\ref{chiral_photospectrum}) suggests that tuning $\etaz=0$ is a second way for closing the gap albeit not at the $\Gamma$ point.   
More precisely, the resulting band structure $E_\az (k) =\pm|\deltaz+\Delta_1 k^2|$
becomes gap-less on the circumference of a circle of radius $k_\az \equiv \sqrt{-\deltaz/\Delta_1}$ 
centered at the $\Gamma$-point in the $k_x$-$k_y$ plane.  
Then, the two bands with the same pseudo-spin cross each other at zero energy into a nodal ring.  However, this $\Gamma$-off gap-closing occurs only for too large values of the parameters (for example, for  $\mathcal{A}^2 \sim 5$~nm$^{-2}, |M_z|>0.8$~eV) and is not present within the desired range of physical parameters shown in Fig.~\ref{phase_diagram}~(a1,b1).
 We also checked that the Chern numbers and pseudo-spin Chern numbers do not change while passing through this kind of gap-closing.

The bands in Fig.~\ref{phase_diagram}(a2,b2) are labeled by their Chern numbers
to show that the phases attributed to the representative states $\chi1$, $\chi2$, $\chi3$ and $\chi4$ are consistent with the Chen numbers.
For example, $(C_+,C_-)=(-1,+1)$ for point $\chi3$ in panel (b2) validates that this is indeed a QPHI.    
The calculated  Chern numbers are verified by the powerful numerical method introduced in Ref.~\cite{Discrete_BZ} by discretizing the Brillouin zone.
We will show that, in addition to the Chern number, two other criteria can be used as the topological invariants. 
The first one is to verify whether the edge modes appear within the band gap of a nanoribbon version of the topological insulator thin film.
The other way would be calculating the Hall conductivity. 
To the sake of completeness, we have conducted both and present the results in the subsequent sections.

\subsection{Edge modes in nanoribbons}\label{sec:ribbon}
If a nanoribbon is cut out of the TI thin film, interesting \emph{edge modes} emerge
which can be used as an alternative criterion for determining distinct phases of diagram. 
Over a square lattice described by lattice vectors $(a\hat x, a \hat y)$,
the photo-induced Hamiltonian, Eq.~(\ref{Photo_Hamiltonian}) is  discretized as a tight-binding Hamiltonian, namely  

\begin{equation}
\label{TB_Hamiltonian}
H=\sum_{\bf r}^{} {\bf c}_{\bf r}^\dagger T_0{\bf c}_{\bf r}+
{\bf c}^\dagger_{\bf r+\hat{x}} T_x{\bf c}_{\bf r}+{\bf c}^\dagger_{\bf r+\hat{y}} T_y{\bf c}_{\bf r}+h.c.
\end{equation} 
where  ${\bf c}_{\bf r}^\dagger$ and ${\bf c}_{\bf r}$ are the well-known creation and annihilation operators of electron at site $\bf{r}$, 
while 
\begin{equation}
\begin{aligned} 
T_0=&\left(\Delta_0^\prime+\frac{4\Delta_1}{a^2}\right)\sigma_z+ m{\alpha}_z\sigma_z,\\
T_x=&\big[-\frac{\Delta_1}{a^2}\sigma_z+i\frac{\hbar v_f}{2a}({\alpha}_z\sigma_y-2A'\Delta_1\sigma_y)\big]e^{ieB_zya/\hbar}\\
T_y=&-\frac{\Delta_1}{a^2}\sigma_z-i\frac{\hbar v_f}{2a}(\sigma_x-2A'\Delta_1{\alpha}_z\sigma_x)\\
\end{aligned}
\end{equation}
are 2$\times$2 onsite-energy ($T_0$) and the hopping-energy matrices ( $T_x$ , $T_y$ ) along the $x$ and $y$  directions, respectively. Here, $\alpha_z$ is pseudo-spin index.
The total mass term is defined by $m = M_z+A'\hbar^2v_f^2$
and the modified dressing-field amplitude is $A'=\mathcal{A}^2/\hbar\Omega=m_{\Omega}/(\hbar v_f)^ 2$.  $B_z$ is the magnitude of perpendicular magnetic field and $y=ja$ is the site position along $\hat{y}$ direction.

Shown in Figs.~\ref{phase_diagram}(a3,b3) is the energy dispersion of Hamiltonian given by Eq.~(\ref{TB_Hamiltonian})
applied to a nanoribbon of width 400~nm in the same four phases $\chi1$ through $\chi4$.  The ribbon is extended along the $x$-direction (i.e. periodic boundary conditions are applied only along the $x$-direction) and thus $k_x$ remains a good quantum number.
The lattice parameter of the square lattice is set to $a=2$~nm throughout the calculations.
Clearly, the edge modes that become present within the gap of the spectra of all four examined phases are \emph{purely pseudo-spin polarized states}. 
In case (i), both QAHI phase chosen from region III  possess two edge modes which are originated purely from the blue bands corresponding to $\az=+$. 
In contrast,  the edge modes of the same phase in case (ii) (purple bands) correspond to $\az=-$. 
Moreover, the edge modes are also classified according to their spatial localization on the top or bottom edges of the nanoribbon,
as labeled by `t' or `b' in Figs.~\ref{phase_diagram}(a3,b3).
Both characters of the edge modes, i.e. the pseudo-spin index and its edge localization, are reversed in the other QAHI phase in region I.
This means that the edge modes are chiral with their pseudo-spin states locked to the ${\bf k}$ direction.
The chirality of the edge mode of regions I and III are opposite.

Intriguingly, 
the QPHI phase $\chi3$, the spectrum of which is depicted in Fig.~\ref{phase_diagram}(b3), contain the edge modes of both pseudo-spins which flow against each other with different velocities, nontheless the difference in velocity of the edge modes is very small which is stemming from the off-resonant regime (low intensity and high-frequency laser field in Eq.\ref{eq:velocity}).
These are indeed {\it helical edge states}.

The situation is schematically illustrated on the bottom panels.

\subsection{Hall Conductivity}\label{S3_ac}
The Chern number is the topological invariant based on which the regions in the phase diagram are separated out.
To validate the obtained Chern numbers, we calculate here the Hall conductivity at zero temperature by means of Kubo formula
\begin{equation}
\begin{aligned}  
G_H= \frac{e^2}{\pi h} \int_{BZ} dk_x dk_y\sum_{E_i<E_F<E_j}
\text{Im}\frac{\langle i|\frac{{\partial H}}{\partial k_x}|j\rangle\langle j|\frac{\partial H}{\partial k_y}|i\rangle}{(E_i-E_j)^2}
\label{Kubo}
\end{aligned}
\end{equation} 
which is proportional to the sum of the Chern numbers of all  occupied bands below the Fermi level as $G_H=Ce^2/h$ when Fermi energy lies inside the gap. When the Fermi energy falls inside a band, the Chern number is not defined but Eq.~\ref{Kubo}  still gives the Hall conductivity.
The integral runs over the Brillouin zone when the Fermi energy lies between the energies of the initial and final states, denoted by $E_i$ and $E_j$, respectively,
where $E_i$ is an eigenvalue of the Hamiltonian Eq.~(\ref{Photo_Hamiltonian}) corresponding to the eigenstate $|i\rangle$. 
The Hall conductivity is calculated for four phases $\chi1$ through $\chi4$ and plotted in Fig.~\ref{kubofig}.
In complete agreement with the Chern numbers,
$G_H$ becomes plateau within the energy gap which is an evidence of the edge modes. 
As a result of scattering, Hall conductivity attributed to the bulk states would be smaller than the edge mode transport. 
The breaking points on the curves are a consequence of participating Van-Hove singularities associated with the conduction and valence band edges in the conductivity. 
In the particular phase $\chi3$, the Hall conductivity is zero inside the band gap which is an evidence for the existence of the NI and QPHI phases. 

It would be worthwile to generate topological states with higher level of Chern numbers leading to justifying the number of the edge states within the gap. Beside of irradiated laser, turning a magnetic field on, conducts us to a fascinating phase diagram with higher number of edge modes appearing within the Landau level gaps.    

\subsection{Landau Level fan diagrams}\label{S4}
By applying a normal magnetic field, two effects emerge: the Zeeman-type coupling and forming Landau levels (LL). 
The Zeeman effect is similar to the exchange interaction between conduction electrons and localized magnetic moments in the TI thin film which we already addressed by introducing $M_z$. 
Although the Zeeman coupling is small compared to the exchange interaction, 
the orbital effects of the perpendicular magnetic field leads to formation of Landau levels with a distinct physics. 
Using the Peierls substitution, the orbital effect is described by modifying the wave vector as
${\bf k} \rightarrow {\boldsymbol \pi}={\bf k}+ e {\bf A}/\hbar$.
The vector potential is in the Landau gauge 
${\bf A}=(yB_z,0,0)$ for a magnetic field applied normal to the thin film plane ${\bf B}= (0,0,-B_z) $. 
The ladder operators acting on the LL orbitals are defined as $a=\ell \pi_+/\sqrt{2}$ and $a^\dagger=\ell \pi_-/\sqrt{2}$ and satisfy $[a,a^\dagger]=1$, where 
$\ell=\sqrt{{\hbar}/{eB_z}}$ is the magnetic length   and $\pi_\pm=\pi_x\pm i\pi_y$.
The harmonic oscillator eigenfunctions represented as $|n\rangle$ step up and down by the ladder operators 
as $a |n \rangle=\sqrt{n} |n-1 \rangle$ and $a^\dagger |n \rangle =\sqrt{n+1} |n+1 \rangle$ with integer quantum number $n=1,2,3, \cdots$. Let us work in the primitive basis set represented as $|u,\uparrow \rangle ,|u,\downarrow \rangle ,|d,\uparrow \rangle ,|d,\downarrow \rangle$. First, the Hamiltonian in the absence
of magnetic field under irradiation of circularly polarized light is deduced from Eq.~(\ref{eq:firsthamil}) as 
\begin{equation}
\begin{aligned}
\centering
& 
{H}(\textbf{k})=\hbar {v}_f (k_y \sigma _x -k_x \sigma _y) \otimes \tau_z +\Delta^\prime (\textbf{k}) \sigma _0 \otimes \tau _x
\\&+V_\text{sia} \sigma _0 \otimes \tau _z
+ m \sigma_z\otimes\tau _0
\\&-2\hbar v_fA'\Delta_1 (k_x \sigma _x+k_y \sigma _y) \otimes \tau_y
\end{aligned}
\label{hamil2}
\end{equation}
where the total mass term includes the exchange field added to the photo-induced mass term; $m=M_z+m_{\Omega}$  and we set $V_\text{sia}=0$. Eventually, switching on the magnetic field $B_z$ results in the  oscillatory Hamiltonian,
\begin{equation}
H'= 
\left(\begin{array}{cccc} 
m & i\omega_1a^\dagger & d_n-\xi/2 & i\omega_1'a^\dagger \\
-i\omega_1a & -m & i\omega_1'a & d_n-\xi/2 \\ 
d_n-\xi/2 & - i\omega_1'a^\dagger & m & -i\omega_1a^\dagger \\
- i\omega_1'a & d_n-\xi/2 & i\omega_1a & -m
\end{array}\right)
\end{equation}
where $d_n=\Delta_0^\prime-\xi a^\dagger a$, $\omega_1=v_f\sqrt{2e B_z \hbar}$, $\omega_1'=2A'\Delta_1\omega_1$ and $\xi=-2e B_z \Delta_1/\hbar$. The eigenvectors for this Hamiltonian is in the form of $(u_1^n |n\rangle, u_2^n |n-1\rangle, u_3^n |n\rangle, u_4^n |n-1\rangle )^T$ 
where $u_i^n$'s are those coefficients which can be determined by diagonalizing the following matrix:
\begin{equation}
H''=
\left(\begin{array}{cccc} 
m & i\omega_n & \tilde{d}_n-\xi/2 & i\omega_n' \\
-i\omega_n & -m & i\omega_n' & \tilde{d}_n-\xi/2 \\ 
\tilde{d}_n-\xi/2 & - i\omega_n' & m & -i\omega_n \\
- i\omega_n' & \tilde{d}_n-\xi/2 & i\omega_n & -m
\end{array}\right)
\label{LL_spectrum}
\end{equation}
in which $\tilde{d}_n=\Delta_0^\prime-\xi n$, $\omega_n=v_f\sqrt{2e B_z \hbar n}$ and $\omega_n'=2A'\Delta_1\omega_n$. The above relations are true for $n\neq 0$. The LL spectrum is calculated by digonalizing of Hamiltonian given in Eq.\ref{LL_spectrum}. In the case of zero mode, $n=0$, the Hamiltonian eigenstates are formed as $(u_1^0 |0\rangle,0 , u_3^0 |0\rangle,0)^T$ with the eigenvalues $E^{\pm}_{0}=m\pm(\Delta_0^\prime-\xi/2)$. 

\begin{figure*}
 \includegraphics[width=\linewidth,trim={0 0 0.25cm 0}, clip]{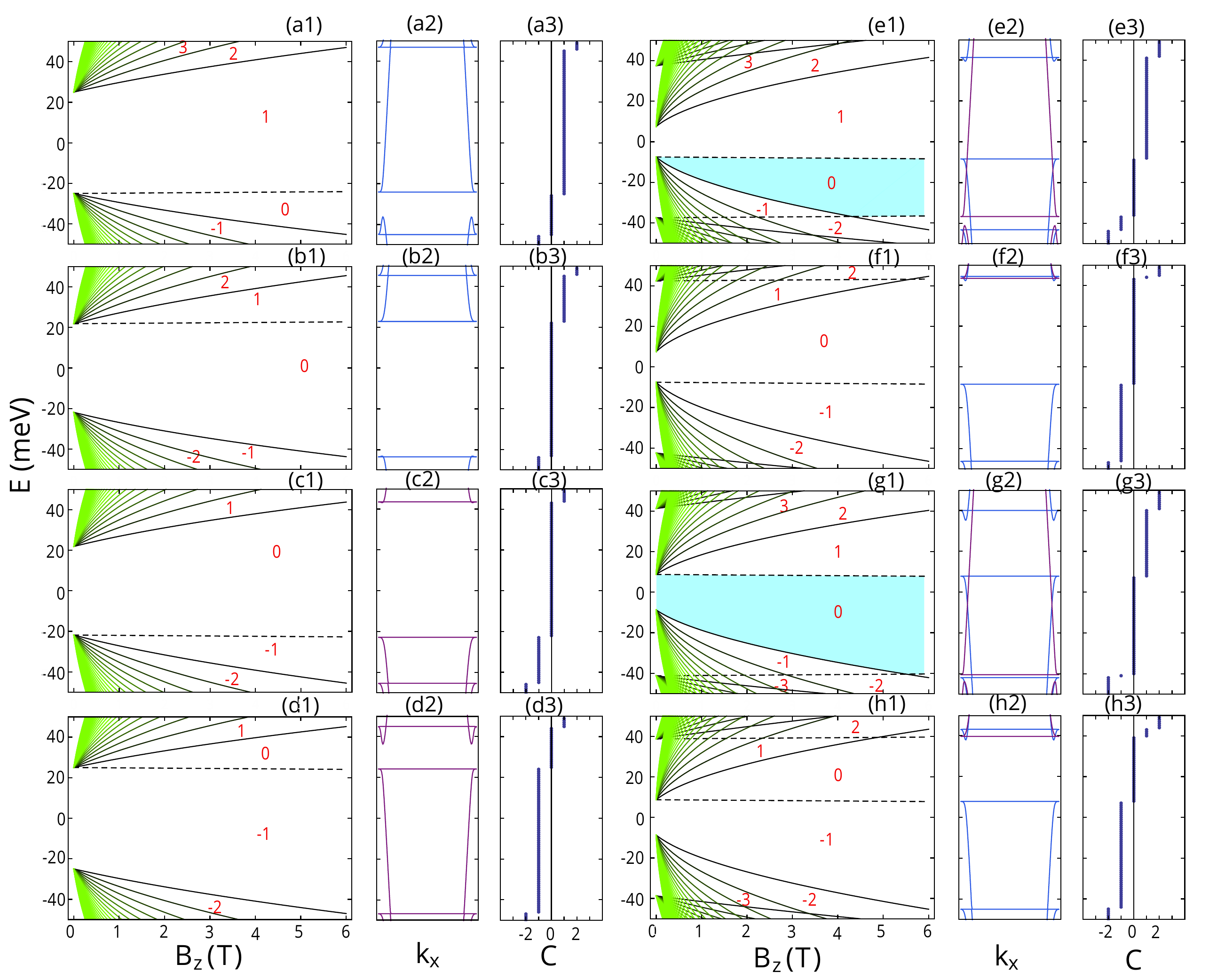}    
\caption{Evolution of the Landau level fan diagram in terms of perpendicular magnetic field for the special states $\chi1$ through $\chi4$ depicted in Fig.\ref{phase_diagram}. In the panels $a1-d1$, we assume $\Delta_0\Delta_1 >0$ while panels $e1-h1$ correspond to the case of $\Delta_0 \Delta_1<0$. Zero modes are indicated by the dashed dark lines while higher Landau levels are displayed with lighter color. The red integers in each region represent the Chern number. Light blue
regions in panels (e1),(g1) correspond to a QPHI phase. The band dispersion of a nanoribbon version of material is plotted in $a2-h2$ at a special magnetic field of $B_z=6$~T where the width and unit cell of length are considered to be 400~nm and 2~nm, respectively. In the dispersions, the blue (purple) bands are attributed to pseudo-spin +1 (-1). To confirm the gained Chern numbers, the Hall conductivity is shown in panels $a3-h3$ as calculated by the Kubo formula on a finite lattice
($200\times10$) in a magnetic field  $B_z=6$~T. The magnitude of  ($M_z,\mathcal{A}^2 Sgn(\Omega)$) are addressed in the caption of Fig. \ref{phase_diagram}. }
\label{Landau}
\end{figure*}

Two zero modes behave linearly with the magnetic field, $E^{\pm}_{0}=m\pm(\Delta_0^\prime+e B_z \Delta_1/\hbar)$ and they cross each other at the critical magnetic field $B^{cr.}_z=-\hbar \Delta_0^\prime/e\Delta_1$ for the case (ii) (${\Delta_0} \Delta_1<0$). These two zero modes have no crossing point in the case of ${\Delta_0} \Delta_1>0$. 
Finally, higher order LLs are calculated for the four specific states $\chi1, \chi2, \chi3$ and $\chi4$  and are plotted in terms of perpendicular magnetic field $B_z$. 
The fan diagrams are shown in Fig.~\ref{Landau} for both case (i) (left) and case (ii) (right). 

Fascinating topological phases with higher Chern numbers (red integer numbers in the LL regions) appear. 
At the zero magnetic field limit, four groups of LL's converge to $E_n=\pm(m\pm{\Delta_0^\prime})$ where zero modes go to $E_0=m\pm{\Delta_0^\prime}$. 
To verify the number of the edge modes appeared in each region of the phase diagram, 
we calculate the Hall conductivity by the Kubo formalism on a finite lattice of size $200 \times 10$ unit cells
at $B_z=6$~T.
An instruction of this method was explained in Ref. \cite{kubofinite}. Periodic boundary conditions are applied by choosing an appropriate length of unit cell. The hopping parameters are those explained in section \ref{sec:ribbon}. The plateaus in the Hall conductivity in Figs.~\ref{Landau}(a3-h3) confirm the Chern numbers assigned to 
the regions depicted in panels a1-h1.
\begin{figure*}   
 \includegraphics[width=0.9\linewidth,trim={0 0 0.25cm 0}, clip]{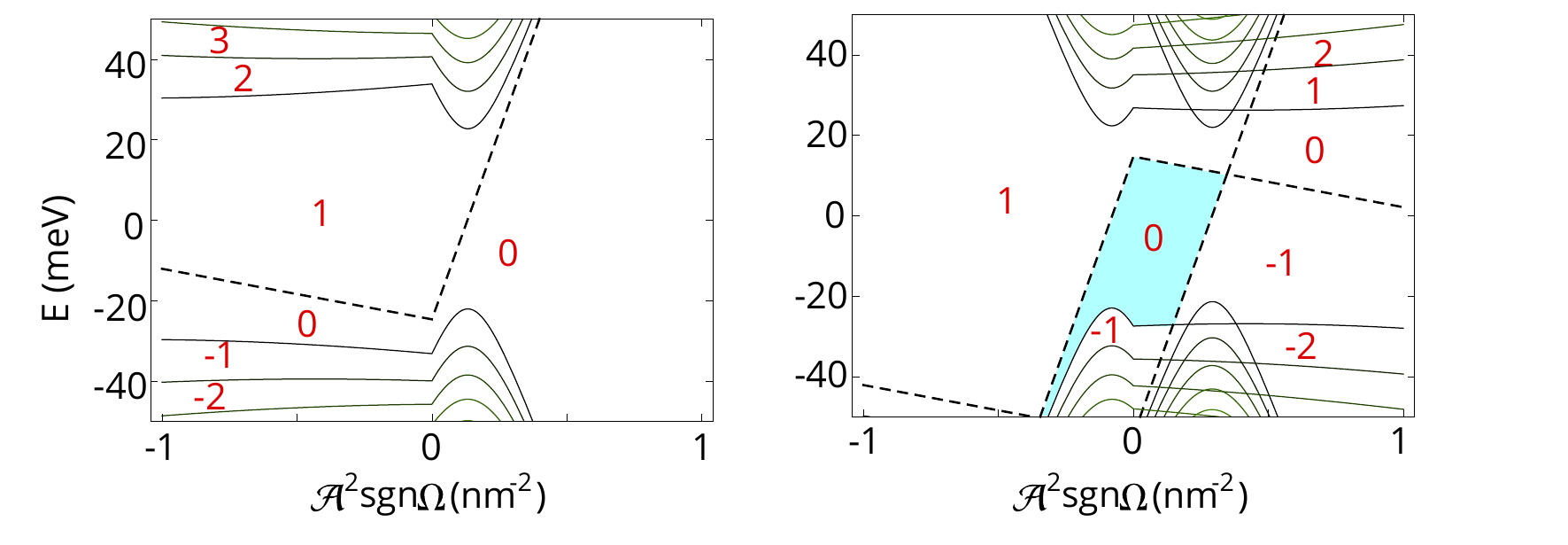}     
 \caption{Landau levels for $M_z=-60meV$(left),$M_z=- 20 meV$(right) and $B_z=2T$ as a function of $\mathcal{A}^2Sgn(\Omega)$ in units of (1/nm$^2$). Panel a(b) is for case i (ii) where ${\Delta_0}{\Delta_1}>0$ (${\Delta_0}{\Delta_1}<0$) respectively. Dashed lines correspond to zero modes. Red integers are attributed to Chern number of each region}
\label{phaseb}
\end{figure*}

\begin{figure*}   
 \includegraphics[width=\linewidth,trim={0 0 0.25cm 0}, clip]{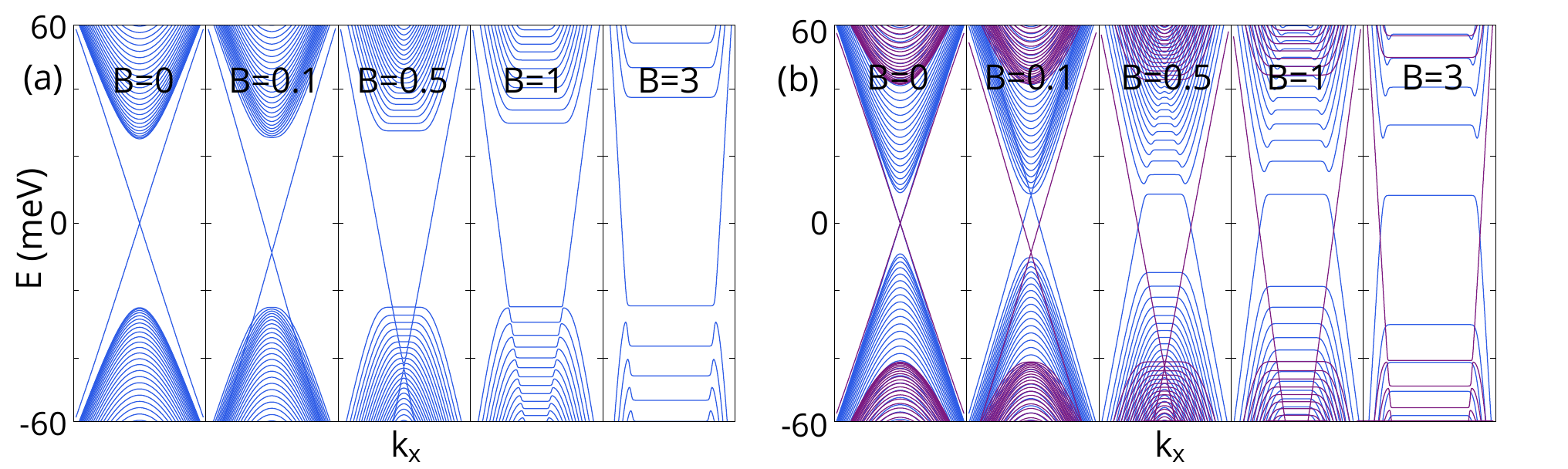}     
 \caption{Evolution of energy bands of the nanoribbon in response to a perpendicular magnetic field. 
 The left panel corresponds to the state $\chi1$ with ${\Delta_0}{\Delta_1}>0$ while the right panel to the state  $\chi3$ with ${\Delta_0}{\Delta_1}<0$ . Blue (purple) color indicates pseudo-spin +1(-1).}
\label{evo}
\end{figure*}

 Furthermore, the energy dispersion of a nanoribbon of width 400~nm and a unit cell of length 2~nm
 subject to a magnetic field $B_z=6$~T is depicted in panels a2-h2, where blue (purple) bands correspond to +1 (-1) pseudo-spin index. The hopping parameters and the method was explained in section \ref{sec:ribbon}. In the presence of magnetic field, based on chosen Landau gauge, the hopping parameter along $x$-direction varies with the $y-$position of sites, and $T_0,T_y$ are the same as before. At finite magnetic field, the dispersion relation of the nanoribbon exibits a periodic structure. We show the result for one period in Figs.~\ref{Landau} and \ref{evo}.
 Emergence of pseudo-spin polarized Landau levels is clear.
 It is notable that because of magnetic field, these modes are localized at the edges of the sample, such that at each Fermi energy one can draw a line and estimate the Chern number by counting the number of intersected energy bands. 
 However, two humps is seen in the band spectra (e.g. the Chern number is zero around $E=-40$~meV in Fig.~\ref{Landau}(a2)) which is originated from the coupling between the LL edge states and also the intrinsic edge states arising from the topological phase of material~\cite{feng}. 
 Although Chern number is zero, two extra edge states attributed to the exotic band with two humps are counterpropagating at one edge of the sample.  In other words, these two clockwise and anti-clockwise edge states manifest themselves in zero Chern number. 
 These two humps are also emerging around $E=38$~meV in Fig.~\ref{Landau}(g2) where Chern number is $+1$. 
 Indeed three edge modes appear at one edge, two of which propagate clockwise and one propagates anti-clockwise.
 
These humps and counterpropagating edge states have been also observed in static systems~\cite{feng,chen2012} and are different from anomalous edge states~\cite{rudner} which occur in the dynamical gaps created by driving field with frequencies lower than the band width. In the high frequency regime considered here, there is no band touching at the dynamical gaps originating from the driving field~\cite{creat_edge}. Applied magnetic field, disturbs effectively Landaue levels falling in the bulk Floquet band.

Figure~\ref{Landau}(a1-h1) implies that at zero Fermi energy, $E_f=0$, 
the phases of neither of the four special points  $\chi1$ through $\chi4$  will change if a very small magnetic field is turned on. 
 The phases remain identical to those depicted in the phase diagram in Fig.~\ref{phase_diagram} at a small magnetic field, namely
$\chi1 (\chi4)$ are QAHI with Chern number +1(-1) while $\chi3$ remains NI (QPHI) for the case i (ii). By increasing the magnetic field, however, new phases appear. 
At other Fermi energies the QAHI phase with negative and positive higher Chern numbers emerges for all $\chi1$ to $\chi4$. 
It is interesting that a new QPHI phase will be also unveiled in a region of the phase diagram (panel e1),
which is indicated by a light-blue color. 
This phase and also QPHI phase in panel g1 will turn into NI if magnetic field exceeds a critical value ($B_z^{cr.}\approx 99,165$~T for points $\chi_1,\chi_3$) when two zero LLs cross each other, nonetheless these  magnitudes of magnetic field are very large and not accessible in experiment. The critical magnetic field is negetive for points $\chi_2,\chi_4$ so there is no QPHI phase in panels (f1,h1) of Fig.~\ref{Landau} .

A detailed exploration of the phase diagram in the presence of $V_{sia}$ is the subject of our future work.  In this work, we focus on the effect of light irradiation in generation of new topological phases by employing magnetic field. In Fig.~\ref{phaseb}, we demonstrate the Landau level dependence on $\mathcal{A}^2 Sgn(\Omega)$ for fixed $M_z=-60$~meV (left) and   $M_z=-20$~meV (right), and magnetic field $B_z=2$~T. 
The zero modes are two straight lines as indicated by dashed lines in Fig.~\ref{phaseb}. 
On the other hand, other modes have an extermum. The phases with zero Chern number are NI in the left panel and can be QPHI (blue region) or NI in the right panel. 
By increasing $\mathcal{A}^2$, zero energy experiences phase transitions such as QAHI~$\rightarrow$~NI in left panel and QAHI~$\rightarrow$~QPHI~$\rightarrow$~QAHI~$\rightarrow$~NI in right panel, which is consistent with phase diagrams \ref{phase_diagram}~(a1,b1) when the Chern number varies as $+1 \rightarrow 0 \rightarrow -1  \rightarrow 0$. 
The phase transitions would be more complicated for other energies.

\subsubsection{Interplay between Landau Level and topological edge modes}
We show in Fig.~\ref{evo} the effect of a magnetic field on the band structure of a nanoribbon.
As before, the ribbon width is 400~nm and the unitcell  periodicity length is 2~nm.
The panels on the left correspond to the case (i),  ${\Delta_0}{\Delta_1}>0$,
and show how the band structure of point $\chi1$ as a potential QAHI evolves with increasing the magnetic field.
The right panels correspond to the case (ii), ${\Delta_0}{\Delta_1}<0$,
and illustrate this evolution for the point $\chi3$ as a candidate for QPHI phase.
By setting $V_{sia}=0$, the band spectra are pseudo-spin polarized as indicted by the blue and purple curves. 
While in the QAHI phase only one band (blue) is inverted, both bands are inverted in the QPHI phase.

A finite magnetic field pushes the band center Dirac cone, which is a representative of topological edge modes, 
down into the valence bands as seen in Fig.~\ref{evo}(a). 
On the other hand, the LL edge modes are gradually formed at stronger magnetic fields such that the bands become $k$-independent.
By coupling of the intrinsic edge modes with LL ones, humps appear in the spectrum at $B=3$~T.
It should be noted that the intrinsic edge modes do not enter into the bulk states but they interact only with the LL edge modes.
More interesting situation will occur for the case (ii) shown in Fig.~\ref{evo}(b). 
In this spectrum, there are two intrinsic edge modes, each attributed to one pseudo-spin state,
which move toward opposite energy sides by applying strong magnetic fields. 
Therefore, the humps for the purple bands are formed in the valence bands while for the blue bands the humps are generated in the conduction bands. 
Moreover, the phase of QAHI (QPHI) in panel a(b) persists in the presence of magnetic field
which is consistent with Fig.~\ref{Landau}a(g).

\subsubsection{Discussion} \label{S3_ba}

\emph{Off-resonant regime:} Based on the Floquet theorem, a non-equilibrium system is projected on a steady-state Hamiltonian giving rise to photon-dressed band spectrum when the material is subject to a time-periodic magnetic field. 
In this work,  we derived an approximate band spectrum with single-photon assisted electrons assuming that there is no optical absorption. Let us call it the ``off-resonant regime'' in which frequency of light is much larger than the band width of the dark material ensuring that there is no photo-transitions between higher photon-assisted Floquet side-bands. 
Instead, by absorption of virtual photons, electrons are excited in the effective modified band structure. textcolor{red}{In this off-resonant regime, there is no anomalous edge states in the bulk Floquet band and the system topology can be described by static AZ classes of effective Hamiltonian\cite{rudner,2013Gomez}.}

\emph{Pseudo-spin conservation:} On the phase diagram, it is supposed that pseudo-spin is conserved. Let us elucidate pseudo-spin conservation in the following~\cite{spinchernnumber,spinchernnumberTI}. QPHI phase generally does not preserve time reversal symmetry. Accompanying this phase with the time reversal symmetry concludes quantum spin Hall insulator which is described completely by the $\mathbb{Z}_2$ invariant~\cite{prl95Z}. 
When time reversal symmetry is not present, the $\mathbb{Z}_2$  invariant would not be well defined and one should use pseudo-spin Chern number to describe this phase. It is well defined even if pseudo-spin is not a conserved quantity~\cite{Prodan,Sheng,spinchernnumber,spinchernnumberTI}. 
The pseudo-spin Chern number is protected by two gaps: band gap and pseudo-spin gap. The latter is the gap in the spectrum of $\mathcal{P}\hat{\alpha}_z\mathcal{P}$ where $\mathcal{P}$ is a projector onto the valence bands. 
So, in order to be able to define pseudo-spin Chern number, there should always exist a gap in the pseudo-spin spectrum. This condition is obviously satisfied in all phases in our case. 

\section{Conclusion}\label{S5}
In this work, we study the emergence of several distinct topological phases in response to illuminating a circularly polarized light on a thin film of topological insulator. 
The modified band structure of the system is constructed by means of photon-dressing electrons in a steady-state. 
In the off-resonant regime, we derived the effective Hamiltonian of TI thin film which is perturbed by one-photon processes in the presence of perpendicular magnetization. 

The irradiated light induces an extra mass term enabling us to  engineer the band gap,
 and as consequence to control the band inversion 
 which is the key for observing non-trivial topological bands. 
 Depending on the thickness of the thin film, as well as the frequency, intensity and polarization of the light, the photo-induced phase diagrams indicate distinct types of topological phases. The phase diagram features unveil a phase transition between QAHI to NI and vice versa for the case $\Delta_0\Delta_1>0$, while for the case $\Delta_0\Delta_1<0$, a QAHI to quantum pseudo-spin Hall insulator (QPHI) and NI transition emerges. 
 The QPHI is a novel phase of the matter in which the edge states are characterized by pseudo-spins instead of electron spins. The pseudo-spin edge states are anisotropic helical states meaning that each of pseudo-spin states have its own velocity. It should be noted that this phase is not as robust against perturbations as QSHI which is protected by time-reversal symmetry. We will go through the effect of structural inversion asymmetry on emergence of light-induced pseudo-spin edge modes in our subsequent work.

All of the assignments on the phase diagram were double checked  by numerically calculating the Chern number of the bands by discretizng the Brillouin zone. Moreover, we calculate the light-induced Hall conductivity by means of Kubo formula. Both sets of calculations confirmed our results. 
For the sake of having a complete view, we deduced the discretized Hamiltonian on a square lattice and inspected the band spectrum of a nanoribbon version of the thin film.
Eventually, the expected edge modes in each region are present in the band gap of these nanoribbons. Furthermore, the current flowing through the top and bottom edges of the nanoribbon were in complete agreement to what we expect of propagating modes in the QPHI and QAHI phases flowing along the nanoribbon edges. 

To manipulate topological materials with higher order of Chern numbers, we derived Landau levels fan diagrams when a perpendicular magnetic field is applied on the TI thin film. The fan diagram demonstrates new phases induced by the magnetic field which can be engineered by varying other thin film parameters as well. Such higher order of Chern numbers is confirmed by Hall conductivity plateaus extracted by the Kubo formula. In the light-induced Landau levels, it is clearly seen that
electron-hole symmetry is broken. To find interplay between Landau level edge modes with topological edge modes coming from light irradiation, we investigate band spectrum of a nanoribbon version of TI thin film in different magnetic fields. It is proved that intrinsic edge modes will be only coupled with the Landau level edge modes and do not interact with the bulk modes. Moreover, some humps appeared because of coupling between these two kinds of edge modes.

\end{document}